\documentstyle[preprint,eqsecnum,aps,epsf]{revtex}
\catcode`\@=11
\def\lsim{\mathrel{\mathpalette\@versim<}}
\def\gsim{\mathrel{\mathpalette\@versim>}}
\def\@versim#1#2{\vcenter{\offinterlineskip
\ialign{$\m@th#1\hfil##\hfil$\crcr#2\crcr\sim\crcr } }}
\catcode`\@=12

\begin{document}

    \tightenlines

\draft
\preprint{KANAZAWA-02-07,    SWAT/02/336}
\title{Scaling Laws and Effective Dimension in Lattice \\
$SU(2)$ Yang-Mills  Theory \\ with a Compactified Extra Dimension
}
\author{ Shinji Ejiri}
\address{Department of Physics, University of Wales Swansea,
Swansea, SA2 8PP, U.K.
}
\author{Shouji Fujimoto and Jisuke Kubo}
\address{
Institute for Theoretical Physics, 
Kanazawa  University, 
Kanazawa 920-1192, Japan
}
\maketitle
\begin{abstract}
Monte Carlo  simulations are performed in a
five-dimensional lattice $SU(2)$ Yang-Mills
theory with a compactified extra dimension,
and scaling laws are studied.
Our simulations indicate that as the compactification radius $R$ decreases,
the confining phase spreads more and more to 
the weak coupling regime, and 
the effective dimension of the theory
changes gradually  from five to four.
Our simulations also indicate that 
the limit $a_4 \to 0$ with
$R/a_4$ kept fixed exists
both in the confining  and deconfining phases
if $R/a_4$ is small enough,
where $a_4$ is the lattice spacing 
in the four-dimensional direction.
 We argue that
the color degrees of freedom in QCD are
confined only for $R < R_{\rm max}$,
where 
a rough estimate shows that
$1/R_{\rm max}$ lies in the TeV range.
Comments on deconstructing
extra  dimensions are given.
\end{abstract}


\section{Introduction}
Since  Kaluza and Klein  \cite{kaluza} found that the 
electromagnetic and gravitational 
 forces can be unified  by introducing
the fifth dimension, their idea has attracted attention for
many decades.
Recently, there have been a lot of  renewed interests in
 field theories with extra  dimensions,
 in which the length scale of 
the extra dimensions 
can be so large that they could  be experimentally observed
\cite{antoniadis1,arkani1,dienes1}.
It is assumed there that for
distances larger than the compactification size,
the massive Kaluza-Klein excitations decouple so that
these theories behave as a  four-dimensional 
continuum theory at low energies.
Since, however, Yang-Mills theories in 
more than four dimensions are   nonrenormalizable,
it is not at all clear that 
the infinite tower of the Kaluza-Klein excitations 
decouple even if 
each massive excitation is suppressed:
A naive expectation of their contribution would be  $\infty \cdot 0$.

At four dimensions, the color degrees of freedom are
confined even for a  weak gauge coupling.
How can a confining four-dimensional Yang-Mills theory emerge
from a higher-dimensional Yang-Mills theory which 
is 
deconfining  in the weak coupling regime  \cite{creutz1,lang,okamoto1} ?
Although the assumption on the 
decoupling of the 
Kaluza-Klein excitations sounds physically correct, it is by no means
trivial that they nonperturbatively  decouple 
in such a way that 
the color confinement takes place even at a  weak gauge coupling.
Recently, we \cite{ejiri1} have started to address related problems
in a concrete example, namely the pure lattice $SU(2)$ Yang-Mills
theory in five dimensions with one 
dimension  compactified on a circle.
We have observed \cite{ejiri1} that
the compactification changes the nature of the phase transition, and
that a second order phase transition,
which does not exist in the uncompactified case,  occurs, 
thus confirming the long-standing expectation  of  Lang, Pilch 
and Skagerstam\cite{lang}.
For the first time, we \cite{ejiri1} have computed
the lattice
$\beta$-function 
in a Yang-Mills theory in 
more than four dimensions, and
have verified nonperturbatively 
the power-law running of the gauge
coupling constant
\cite{dienes1,peskin,veneziano,kobayashi}.

In this paper we would like to extend the analyses of 
\cite{ejiri1}. We first observe that
if the compactification radius becomes 
smaller and smaller, 
the confining phase spreads more and more to 
the weak coupling regime.
At the same time we  compute the effective dimension
\cite{connor,kubo2}, and see
that the theory behaves more and more as a 
four-dimensional Yang-Mills theory.
Based on this result,
we argue that
the color degrees of freedom in QCD are
confined only for $R < R_{\rm max}$.
Our 
rough estimate yields that
$1/R_{\rm max}$ lies in the TeV range.

Our calculations of the potential between
two static quarks separated in the four-dimensional
subspace show that 
the deconfining phase is  a Coulomb phase.
We then discuss
the nature of the transition from
the deconfining phase  to the confining phase
for fixed values of  $R/a_4$,
where $a_4$ is the lattice spacing 
in the four-dimensional direction.
We
confirm that if $R/a_4$ is small enough, it
is consistent with 
a  second order transition \footnote{
The phase we are talking about in this paper
is defined by the potential between two quarks
separated in the {\em four-dimensional } subspace.
This definition should not be confused with the 
definition by
the Ployakov
loop that extends into the fifth dimension.
In the case of the phase transition
measured by the Ployakov loop, too,
the change from  first  to second order 
occurs \cite{ejiri1} as we have already mentioned.}.
Combined with the result of \cite{ejiri1},
we  therefore come to the conclusion
that as we decrease the value of $R\Lambda$,
the first order transition for  large
values of $R\Lambda$ changes to 
a cross over transition, and finally
becomes of second order.

We give some
nonperturbative comments on deconstructing
extra dimensions \cite{arkani2} 
in Conclusion.

\section{Effective dimension}
In order to take into account  the  compactification effects  in this theory,
it is crucial to use  anisotropic
lattices  \cite{burgers} which has different lattice spacings, $a_4$ and 
$a_5$,
in the four-dimensional directions and in the fifth direction.
For definiteness we  employ
the Wilson action for pure $SU(2)$ lattice gauge theory
\begin{eqnarray}
S &=& \frac{\beta}{\gamma}~\sum_{P_4}~[~
1-\frac{1}{2}~\mbox{Re Tr} ~U_{P_4}~]+
\beta \gamma~\sum_{P_5}~[~
1-\frac{1}{2}~\mbox{Re Tr}~ U_{P_5}~]~,
\label{action}
\end{eqnarray}
where $U_{P_4}$ denote  plaquette variables
in the four-dimensional sublattice,
 and $U_{P_5}$ are those which are extended in the
fifth dimension.
The gauge coupling constant $g_5$ has the dimension
of $\sqrt{a_4}$, which is related to $ \beta $ by 
\begin{eqnarray}
a_4 g_{5}^{-2} = \beta/4
\label{beta}
\end{eqnarray}
at the tree level.
Periodic boundary conditions are imposed in all  directions,
and we use the lattice size of the form $N_4^4\times N_5$
(we mostly use $N_4=12$ and $N_5=4$).
The compactification radius is defined as $R=a_5 N_5/2\pi$
if $N_4 a_4 > N_5 a_5$ is satisfied,  and 
the 
correlation-anisotropy parameter is defined as
$ \xi =a_4/a_5$.
The tree level relation $\gamma =\xi$  will be
modified at the quantum level \cite{burgers}, and
throughout this paper we assume that
the $\xi-\gamma$ relation obtained 
in \cite{ejiri1} 
is satisfied both in the confining and deconfining phases.
Simulations are performed for 
\begin{eqnarray}
\gamma = 3.6, 4.0, 4.6, 5.0~,
\end{eqnarray}
which is equivalent to \cite{ejiri1}
\begin{eqnarray}
\frac{2\pi R}{a_4} &=&\frac{N_5 a_5}{a_4} =\frac{4}{\xi} 
\approx  0.72, 0.64,  0.55, 0.50~,
\label{Rovera}
\end{eqnarray}
where we have used $N_5=4$ above.
We have chosen this range of $2\pi R/a_4$,
because we expect from the previous
calculations \cite{ejiri1} that
the limit $2\pi R/a_4 \to 0$ may exist and
we observe some scaling behavior.

To define the physical scale in the confining phase, 
we use the string tension $\sigma$ between two static quarks that
are separated in the four-dimensional subspace.
Since  the string tension is a physical quantity, 
the lattice string tension
$\sigma_{L}$ should behave like $a_4^2$ as $a_4 \to 0$, where
$a_4$ can be related by  the $\beta$ function to 
the dimensionless bare gauge coupling  
\begin{eqnarray}
g^2 &=&\frac{8\pi}{\beta}=\Lambda g_5^2~,
\label{g}
\end{eqnarray}
where we have identified $\Lambda$ with $2\pi/a_4$ because
$4\times (\pi/a_4)^2=(2\pi/a_4)^2$.
Since we expect that the massive Kaluza-Klein excitations
decouple increasingly as $2\pi R/a_4$ decreases, 
the lattice $\beta$ function $\beta_L$ cannot assume
a purely  five- or  four-dimensional form.
Instead, we expect a continuous change of its form.
This is quantitatively expressed by the so-called
effective dimension $D_{\rm eff}$ which is
a function of  $2\pi R/a_4$ \cite{kubo2}. 
So, we assume that $\beta_L$ can be written as
\begin{eqnarray}
\beta_L &=&-a_4\frac{d g^2}{d a_4}=
[~D_{\rm eff}(2\pi R/a_4)-4~]~g^2-\frac{2b}{16 \pi^2}g^4~,
\label{betaL1}
\end{eqnarray}
where $b=22/3-2/3=20/3$.  Therefore,  the evolution equation 
of $g^2$ can be easily integrated in the case that $2\pi R/a_4$ is 
kept fixed while $a_4$ changes.
We obtain for this case
\begin{eqnarray}
\sqrt{\sigma_{L}} &\sim& a_4 \sim 
\left(\frac{2b}{16\pi^2} \frac{1}{D_{\rm eff}-4} 
-\frac{\beta}{8\pi}\right)^{1/(D_{\rm eff}-4)}~.
\label{deff}
\end{eqnarray}
It is important to notice 
that as  $D_{\rm eff} \to 4$, we obtain the logarithmic form
\begin{eqnarray}
g^{-2} &=&\beta/8\pi=(2b/16\pi^2)\ln a_4 +const.
\label{log}
\end{eqnarray}
That is, if we can show that
the effective dimension $D_{\rm eff}$ in the confining phase
varies from $5$ to $4$ as
$2\pi R/a_4$ decreases from a larger value to a smaller value,
we show the continuous decoupling of the Kaluza-Klein excitations, and
the confining phase spreads more and more to the weak coupling regime as 
$R$ decreases.

\section{Confining phase}
Now we come to the results of our Monte Carlo simulations
on a $12^4 \times 4$ lattice.
We use  the Creutz ratio $\chi(I,J)$ obtained from the 
rectangular Wilson loops
$W(I,J)$  with lengths of $I$ and 
$J$ in the four-dimensional subspace.
We assume that the Creutz ratio takes the form
\begin{eqnarray}
\chi(I,J)=\chi_0-\chi_1\left( \frac{1}{I(I-1)}+\frac{1}{J(J-1)} \right)
+\chi_2\left( \frac{1}{I(I-1)J(J-1)} \right)~,
\label{chi}
\end{eqnarray}
and we identity  $\chi_0$ 
with the lattice  string tension $\sigma_{L}$.
We have generated 2500 configurations for each simulation point 
after thermalization, and 
the Wilson loops are measured every 5 configurations for the calculation 
of a Creutz ratio.
Errors are estimated by the jackknife method.
The filled symbols in Fig.~1 are the result
obtained from the Monte Carlo simulations 
with $\gamma=5.0$,
where 
the vertical axis  stands 
for $\sqrt{\chi_0} = \sqrt{\sigma_{L}}$ and the horizontal axis stands for
$\beta $. 
We have also calculated $\sigma_{L}$ from
the static potential to make it sure that
$\sigma_{L}$ obtained from the Creutz ratios
is reliable \footnote{
We give more details
of calculating the static potential in section V
when calculating the potential in the Coulomb phase.}.
The static potential we have assumed has the
form
\begin{eqnarray}
V(X)=C_0-C_1\frac{1}{X}+C_2\left(\frac{1}{X}-\left[\frac{1}{X}\right
]\right)
+C_3X~,
\label{st-pot}
\end{eqnarray}
where $[1/X]$ is the  three-dimensional Coulomb potential on a 
lattice, and is given by
\begin{eqnarray}
\left[\frac{1}{X}\right]=
4\pi\int_{-\pi}^{\pi}\frac{d^3p}{(2\pi)^3}
\frac{\exp\{i\sum_iX_i\sin(p_i/2)\}}{\sum_{i=1}^{4}\sin^2(p_i/2)}~.
\label{3dcoulomb}
\end{eqnarray}
The open symbols in Fig.~1 correspond to
the result obtained from the static potential.
Comparing two results in 
Fig.~1 we see that the lattice string tensions obtained from the Creutz ratios
agree with those obtained from the static potential.
We have made the same comparison
for different values of $\gamma$, and found the same result.
So, in the following analyses we use
only the lattice string tensions from the Creutz ratio,
because we  have more data for this case 
and we do not
want to mix  data obtained by two different
methods.

We see from Fig.~1 that above $\beta \gsim 3.0$ 
the square root of
the lattice string tension $\sqrt{\sigma_{L}}$  decreases first linearly until 
$\beta \sim 3.3$, and then its slope becomes milder.
The tail for large $\beta$ is certainly due to the finite
lattice size effects, but the change from the linear decrease of $\sqrt{\sigma_{L}}$
to a milder one around $\beta \sim 3.3$ 
may indicate that the theoretical expectation  (\ref{deff}) is correct.
Although it is in principle possible to check by increasing the
lattice size how much finite lattice size effects
may be contained in the tail of $\sqrt{\sigma_{L}}$, it is 
impossible to do this at the moment because of 
the limitation of the computing facility given to us.
Below we sketch how we confirm Eq. (\ref{deff}) and compute
$D_{\rm eff}$.

The effective dimension 
can be obtained by fitting the function (\ref{deff}) to the data.
To this end, we first choose four neighboring data points
that lie around the middle of the data set for a given $\gamma$, 
and using these points, we fit the function (\ref{deff}) to
obtain the effective dimension.
(In the case for $\gamma=5.0$ for instance,
we use the data points at $\beta=3.20, 3.22,3.24$ and $3.26$.)
Then we increase the number
of the data points, to be used, by two by including
the next neighboring data point in both sides. In doing so, we obtain 
the effective dimension and also $\chi^2/\mbox{DOF}$ as 
a function of the number  $n$ of the data points that 
are used for  a fit.
We repeat the same analysis for the different values of $2\pi R/a_4$
given in (\ref{Rovera}). The results are  shown in Fig.~2, 3 and
in Table I.
In Fig.~3,  the vertical axis stand for 
$(D_{\rm eff}-4)^{-1}$ and the error bar is computed from
 $\chi^2/\mbox{DOF}$.
We see that as $n$ increases, the error bar  decreases 
and the central values converge. The results are
summarized in Table I, and
we see that 
the effective dimension $D_{\rm eff}$ decreases gradually  
from $4.7057(55)$ to
$4.5230(82)$ as $\gamma$ increases 
from $3.6$ to $5.0$, which means,
as $2\pi R/a_4$ decreases from $0.72$ to $0.5$ (see (\ref{Rovera})).

The $\beta^*$  in Table I is the value at which
$\sigma_L$ and hence $a_4$ should vanish 
if the theoretical assumption (\ref{deff}) is correct and extrapolated for 
larger values of $\beta$ (see also Fig.~2).
We  emphasize that our results indicate that 
the limit $a_4 \to 0$ with
$R/a_4$ kept fixed exists
 in the confining phase
at finite $\beta$.

\section{The maximal radius}
The same analysis in the real QCD in section III would 
constrain the size of the compactification radius in QCD,
which we would like to estimate without  detailed
calculations.  
To see that there exists the maximal radius 
for color confinement in the four-dimensional subspace, we 
recall the results obtained in the previous section and those from the next 
section:
\begin{eqnarray}
g^2 \gsim (\lsim )  (D_{\rm eff}-4)(16\pi^2/2b)~~
\mbox{for the (de)confining phase}~.
\label{condition}
\end{eqnarray}
Therefore, for a given value of $D_{\rm eff}$,
there should exist the smallest value of $g^2$
for  color confinement to occur,
which is $\sim (D_{\rm eff}-4)(16\pi^2/2b)$.
The question is how $g^2$ can be related
to the gauge coupling $g_{\rm kk}^2$
of the Kaluza-Klein theory, the four-dimensional theory
with the Kaluza-Klein tower.
At the tree level, it is  $g^2_{\rm kk}=g^2 (2\pi R \Lambda)^{-1}$,
but in higher orders this relation 
will receive quantum corrections, where we have used
$\Lambda^2=(\pi/a_4)^2\times 4$.
To answer the question, 
we first assume that $D_{\rm eff}(R\Lambda) \to 4 (5)$
as $R\Lambda \to 0 (\infty)$, and 
we consider a
redefinition of $g^2$ according to \cite{connor,kubo2}
\begin{eqnarray}
g^2_{\rm kk}&=& \eta^{-1} (R \Lambda) g^2~,~
\eta (t) = \exp \int_0^t \frac{dt'}{t'} ( D_{\rm eff}(t')-4)~.
\label{gkk}
\end{eqnarray}
Note that the $\beta$ function of $g^2_{\rm kk}$ becomes
\begin{eqnarray}
\beta_{\rm kk} &=& -(2 b/16\pi^2)\eta (R \Lambda) g^4_{\rm kk}~.
\end{eqnarray}
Since the function $\eta (R \Lambda)$
becomes proportional to $R\Lambda$ as 
$R \Lambda \to \infty$ 
\footnote{The proportionality constant depends on 
$D_{\rm eff}$ as a function of $t$, which, however, 
depends on
the regularization used \cite{kubo2}. Therefore, 
the lattice regularization
does not reproduce the same coefficient \cite{ejiri1} obtained in
\cite{dienes1}.},   
the new
gauge coupling
 describes the power-law behavior \cite{dienes1,peskin,veneziano,kobayashi}.
Furthermore,
 we see  from (\ref{gkk}) that
$g^2_{\rm kk}$ approaches  $g^2$ as $R\Lambda $ approaches $ 0$.
Recalling now the assumption
that $D_{\rm eff}$ approaches $4$ as  $R\Lambda$ approaches $ 0$
and Eq. (\ref{log}),
we see that the renormalization group flow of
the new gauge coupling $g^2_{\rm kk}$ for  small $R\Lambda$
takes exactly the same form as the one for the
effective, four-dimensional 
theory without the Kaluza-Klein tower.
Therefore, we assume that $g_{\rm kk}^2$ is
the gauge coupling 
of the four-dimensional theory
with the Kaluza-Klein tower.

Now, suppose that QCD results from
a five-dimensional QCD.
As we have argued above,
 $g^2$ becomes $g_{\rm kk}^2$
at low energies, and we
then identify $2 \pi/a_4$ with
the physical scale $\Lambda$  of the effective theory,
rather than with the ultraviolet cutoff.
Since  $g^2 (M_Z) /4\pi \simeq 0.12$
in QCD and $b=7$, the constraint (\ref{condition})
can be converted to that of  the effective dimension, i.e.,
$D_{\rm eff}(R M_Z) \lsim 4.13$.
Therefore, if we know the function $D_{\rm eff}(t)$
exactly, we can calculate the range of $t$ for which 
the inequality  (\ref{condition}) is satisfied.
From the  result given in Table I we  find that 
the effective dimension as a function of $t$
can be written as $D_{\rm eff}(t)\simeq
4+t$. Assuming that this function
can be used even for small $t$, we then obtain 
$R\Lambda \lsim 0.13$,  which implies that
\begin{eqnarray}
1/R \gsim O(1) \mbox{ TeV}
\label{rmax}
\end{eqnarray}
should be satisfied for the
color degrees of freedom in QCD to be confined.
The number above should not be taken very seriously,
because
our estimate is based on many
theoretical assumptions, which can be justified 
if we perform simulations on five-dimensional, compactified
$SU(3)$ lattice gauge theory.
The crucial point is that
there exists the maximal radius.

\section{Coulomb phase}
The confining phase 
shrinks as $R$ decreases, which we have already seen above.
Next we would like to show that  the deconfining phase is 
 a Coulomb phase.
To begin with, we consider the 
Wilson loop $W(\vec{x},t)$ 
at the tree level  in continuum perturbation theory.
The static potential can be obtained by
\begin{eqnarray}
V(x) &=& \lim_{t \to \infty} (\ln W(\vec{x},t))/t =
-\frac{3}{4} g_5^2~\frac{1}{2\pi R}\frac{1}{4\pi x}\coth\left(\frac{x}{2R}\right)
\nonumber\\
& = &-\frac{3}{4} g_5^2~ \left\{
\begin{array}{@{\,}ll}
    \frac{\displaystyle 1}{\displaystyle 4\pi^2x^2} &
    \left(\frac{\displaystyle x}{\displaystyle 2R}\ll 1\right)\\
    & \\
    \frac{\displaystyle 1}{\displaystyle 2\pi R}\frac{\displaystyle
    1}{\displaystyle 4\pi x} & \left(\frac{\displaystyle x}{\displaystyle 2R}
    \gg 1\right)\\
  \end{array}~.
\right.
\label{pot0}
\end{eqnarray}
We have the  usual  Coulomb potential for $ x/2R\gg 1$, and we see that
the dimensionless gauge coupling $\hat{g}$, normalized for
the four-dimensional Yang-Mills theory  at the tree level, is given by
$\hat{g}=g_5/\sqrt{2\pi R}$ as well known 
\cite{arkani1,dienes1}. 
The corresponding expression on a lattice is
\begin{eqnarray}
V_L(X) &=& \lim_{T\to\infty}\ln W(X,T)/W(X,T+1)~,
\label{VL}
\end{eqnarray}
where $W(X,T)$ is a lattice Wilson loop.
The lattice distances $X$ and $T$ are made dimensionless by 
dividing by $a_4$.
We are interested in
the potential between two static quarks that are separated
in four dimensions, and therefore, $X$ and $T$ are supposed to
be in the four-dimensional  sublattice.
Since in the actual calculations we cannot take the limit $T \to \infty$,
we  consider also off-axis loops and use
the standard smearing techniques \cite{ape} to improve
the convergence of approximants with increasing $T$.
Our smearing procedure consists of iteratively replacing
each spatial (three-dimensional) link by the sum of itself and 
the neighboring four
spatial staples with a weight parameter $\epsilon$,
\begin{eqnarray}
U_i(x,y)&\to& U'_i(x,y)=
{\cal P}_{SU(2)}\left(
  U_i(x,y)+\epsilon\sum_{j(\ne i)=1}^3F_{ij}(x,y)
\right)
\label{smear}\\
F_{ij}(x,y)&=&U_j(x,y)U_i(x+\hat{j},y)U^\dagger_j(x+\hat{i},y)
+U^\dagger_j(x-\hat{i},y)U_i(x-\hat{i},y)U_j(x-\hat{i}+\hat{j},y)~,
\nonumber
\end{eqnarray}
where
${\cal P}_{SU(2)}$ denotes a projection operator, back onto the $SU(2)$
manifold.

We have generated 10000 configurations for each simulation point after
thermalization, and the smeared Wilson loops are measured every 100
configurations for the calculation of the static potential.
We iterate Eq.(\ref{smear}) 60 times with $\epsilon=0.1$
in the case of the confining phase,
100 times with $\epsilon=0.2$ in the case of the Coulomb phase.
In Fig.~4 we show the result (filled symbols)
for  the lattice potential 
$V_L(X)$ as a function of $X$
at $\beta=5.0$ and $\gamma=5.0$ (which is equivalent to $2\pi R/a_4=0.5)$.
The condition $ x/2R\gg 1$ to obtain 
a $1/X$ potential becomes $X \gg 1/2\pi$ in this case, and 
we assume that the lattice potential $V_L(X)$ takes the form
\begin{eqnarray}
V_L(X)=C_0-C_1\frac{1}{X}+C_2\left(\frac{1}{X}-\left[\frac{1}{X}\right
]\right)~,
\label{pot}
\end{eqnarray}
where is $[1/X]$  (the  three-dimensional Coulomb potential  on a 
lattice is given in (\ref{3dcoulomb}).
The first term of (\ref{pot}) is the unphysical self energy,
the second term is a rotationally invariant part of the Coulomb potential,
and third term is the most dominant part of its breaking.
From a $\chi^2$ fit we find that $C_0=0.3230(14),
C_1=0.1086(30),C_2=0.0776(27)$. The fitted lattice potential with the $C_2$
term in (\ref{pot}) suppressed, i.e.
\begin{eqnarray}
V(X)=C_0-C_1\frac{1}{X}~,
\label{pot1}
\end{eqnarray}
is the dotted curve in 
Fig.~4, 
while the open symbols stand for the 
rotationally invariant data points. 
We see that the data justify the assumption 
that the deconfining phase is  a Coulomb phase.

\section{Nature of the phase transition}
As the next task we consider the nature of the transition
from the confining phase to  the Coulomb phase.
In the confining phase our data indicate
that 
the limit $a_4 \to 0$ with
$R/a_4$ kept fixed exists at finite $\beta$.
If we can show that $a_4$ also vanishes
at the same value of $\beta$ in the Coulomb phase,
the transition from the confining phase to  the Coulomb phase
is of second order.

To this end, we have to define the scale in  the Coulomb phase.
In the naive continuum theory there are two dimensional quantities,
the gauge coupling $g_5$ and the compactification radius $R$.
Therefore,
we assume that $R$ and
the low-energy value of $g_5$ 
are independent physical quantities at the quantum level, too.
We then consider the limit $a_4 \to 0$
with $2\pi R/a_4$ kept fixed, which is the same limiting process
we have considered in the confining phase.
In this limit, the quantity 
$g_5^2/2\pi R$ (the coefficient $C_1$ of the 
tree-level Coulomb potential (\ref{pot})) has to
diverge because $R \to 0$ while $g_5$ should remain finite by assumption.
So naively one expects the scaling law
$C_1^{-1}\sim R \sim a_4 \sim (\beta-\beta^*)$,
where $\beta^*$ is
the critical value of $\beta$ 
at which $\sigma_L^{1/2} \sim a_4$ 
 vanishes.
In Fig.~5 we plot $C_1^{-1}$ versus $\beta$ for different values 
of $\gamma$ (or $2\pi R/a_4$ of (\ref{Rovera})).
We see that $C_1^{-1}$ linearly decreases, 
and make therefore  a theoretical ansatz for the scaling law:
\begin{eqnarray}
C_1^{-1} &=& D_0-D_1 \beta~.
\label{c1fit}
\end{eqnarray}
For $\gamma=4.6$, for instance, 
a $\chi^2$ fit yields that $D_0=9.16(36)$ and 
$D_1=3.894(77)$.
If the tree-level equation (\ref{pot0}) would be 
correct at the quantum level, too, then it would mean that
$a_4$ vanishes at $\beta=D_0/D_1=2.35(14)$ in the deconfining phase.
This would contradict the assumption that
in the confining phase the lattice spacing $a_4$ approaches zero
as $\beta$ approaches $ \sim 3.7 $ for $\gamma =4.6$ (see Table I).
This does not necessarily mean that the transition
from the deconfining phase to the confining 
one is a first order transition or  a cross over transition.
It may be well possible that the tree-level form (\ref{pot0})
receives  quantum corrections in such a way
that the transition is indeed of second order.
Therefore, we consider
possible quantum corrections to
$C_1^{-1}$ which are 
consistent with the
scaling law in Fig.~5 and the value of $\beta^*$ in the
confining phase (given Table I).
Since $C_1^{-1}$ being dimensionless can depend only on
the combination $R/g^2_5$, the correction   can only be a 
constant, i.e.,
\begin{eqnarray}
C_1^{-1} &\sim & 2\pi R/g^2_5+\alpha~\mbox{or}~~
C_1 \sim  \frac{g^2_5}{2 \pi R+\alpha g_5^2}~.
\label{c1}
\end{eqnarray}
In Table III we give the results of the fits, from which we 
find that the ansatz for the 
nonperturbative quantum correction to the coefficient of
the Coulomb potential (\ref{pot}) is consistent with our data,
and we conclude that
\begin{eqnarray}
\alpha &=&5.1\pm 0.7~,
\label{alpha}
\end{eqnarray}
where we have not included the data for $\gamma=5.0$ in
(\ref{alpha}), because the error for this case is
very large compared with others.
This indicates that the assumption that 
the transition from
the confining to the deconfining phase is a second order transition
is consistent with the data.
Note that the transition for small values of $\gamma$, 
or large values of $R\Lambda$, is of first order
\cite{lang,ejiri1}.
We expect that the first order transition for  large
values of $R\Lambda$ changes to 
a cross over transition, and finally
to the second order
transition as we decrease the value of $R\Lambda$ \footnote{
In the case of the phase transition
measured by the Ployakov
loop that extends into the fifth dimension,
the change from the first to second order 
happens at a certain value of $\gamma$ \cite{ejiri1}.}

The nonperturbative correction (\ref{c1}) means that
the tree level relation $\hat{g}^2=g_5^2/2\pi R$ should be 
modified to
\begin{eqnarray}
\hat{g}^2 &=& \frac{g_5^2}{2\pi R}(1+\alpha \frac{g_5^2}{2\pi R})^{-1}~.
\label{matching}
\end{eqnarray}
Since $\alpha$ is large, the correction is not small.
The Coulomb phase may be of  phenomenological
importance, because the color degrees of freedom
do not need to be always confined. The $SU(2)$ part of the standard
model, for instance,  could
result from  a higher-dimensional Yang-Mills theory
in the Coulomb phase. Then the equation like Eq. (\ref{matching}) 
defines the matching condition.

\section{Conclusion}
In this paper we have performed Monte Carlo  simulations in a
five-dimensional lattice $SU(2)$ Yang-Mills
theory, where  we have compactified one extra dimension.
We have found that as the compactification radius $R$ decreases,
the confining phase spreads more and more to 
the weak coupling regime, and 
the effective dimension of the theory gradually
changes from five to four.
Our data indicate that for fixed $R/a_4$ the transition from 
the deconfining phase to 
 the Coulomb phase  is of second order
 if $R/a_4$ is small enough.

Assuming that the real
four-dimensional QCD results from
the five-dimensional QCD at low-energies, 
we have estimated the largest compactification
radius $R_{\rm max}$ so that the color degrees of freedom 
in four dimensions are confined.  
Our estimate (\ref{rmax}) should be understood as the
first try, because our calculations are based on many
theoretical assumptions, which can be justified 
if we perform simulations on five-dimensional, compactified
$SU(3)$ lattice gauge theory.
The striking fact is that
there exists the maximal radius, and this
may give an important phenomenological constraint
for model building based on the  Kaluza-Klein theories.

The parameter regime we have considered
in the present work corresponds to the regime 
in which 
the Kaluza-Klein idea is expected to be realized:
At short distances we have the five-dimensional 
rotational invariance, and at long distances,
the Kaluza-Klein excitations decouple so that
the low-energy effective theory is a four-dimensional Yang-Mills theory.
We found no indication that would contradict this picture.
Moreover, the compactified five-dimensional theory,
which is perturbatively nonrenormalizable,  
has the predictive power  (unless examined at very short distances),
which we conclude from the 
scaling laws we observe. 
(The readers are also invited to \cite{kubo3}.)

The parameter regime that corresponds to
deconstructing  extra dimensions \cite{arkani2} is not
the same as above  \cite{murata}; two phases are nonperturbatively 
separated  \cite{ejiri1,murata}.
In the phase for the conventional Kaluza-Klein theory,
the vacuum expectation value of the Ployakov
loop (that extends into the fifth dimension) is nonzero \cite{ejiri1}, while
it vanishes \cite{murata} in the phase for deconstructing  extra dimensions.
(The phase for deconstructing  extra dimensions
is the one in which 
the layer structure in five-dimensional gauge theories
can be realized  \cite{fu}).
Although it is not at al clear that
the five-dimensional 
rotational invariance
at short distances is recovered,
it looks at the moment as if
two different confining four-dimensional Yang-Mills theories
could result from two different phases (one from each) of a
five-dimensional theory. 
The difference is purely nonperturbative.
It will be very exiting to investigate this
difference more in detail,
especially, in supersymmetric cases,
where one has already analytic results,
and it is shown that the five-dimensional
Lorentz invariance
is recovered \cite{csaki}.

\acknowledgments

This work is supported by the Grants-in-Aid
for Scientific Research  from 
 the Japan Society for the Promotion of Science (JSPS) (No. 11640266,
 No. 13135210).
We would like to thank for useful discussions
K-I. Aoki, V. Bornyakov, M. Murata, 
H. Nakano, 
M. Polikarpov, G. Schierholz, H. So,  T. Suzuki and H. Terao.

\begin{figure}
\epsfxsize=12cm
\centerline{\epsfbox{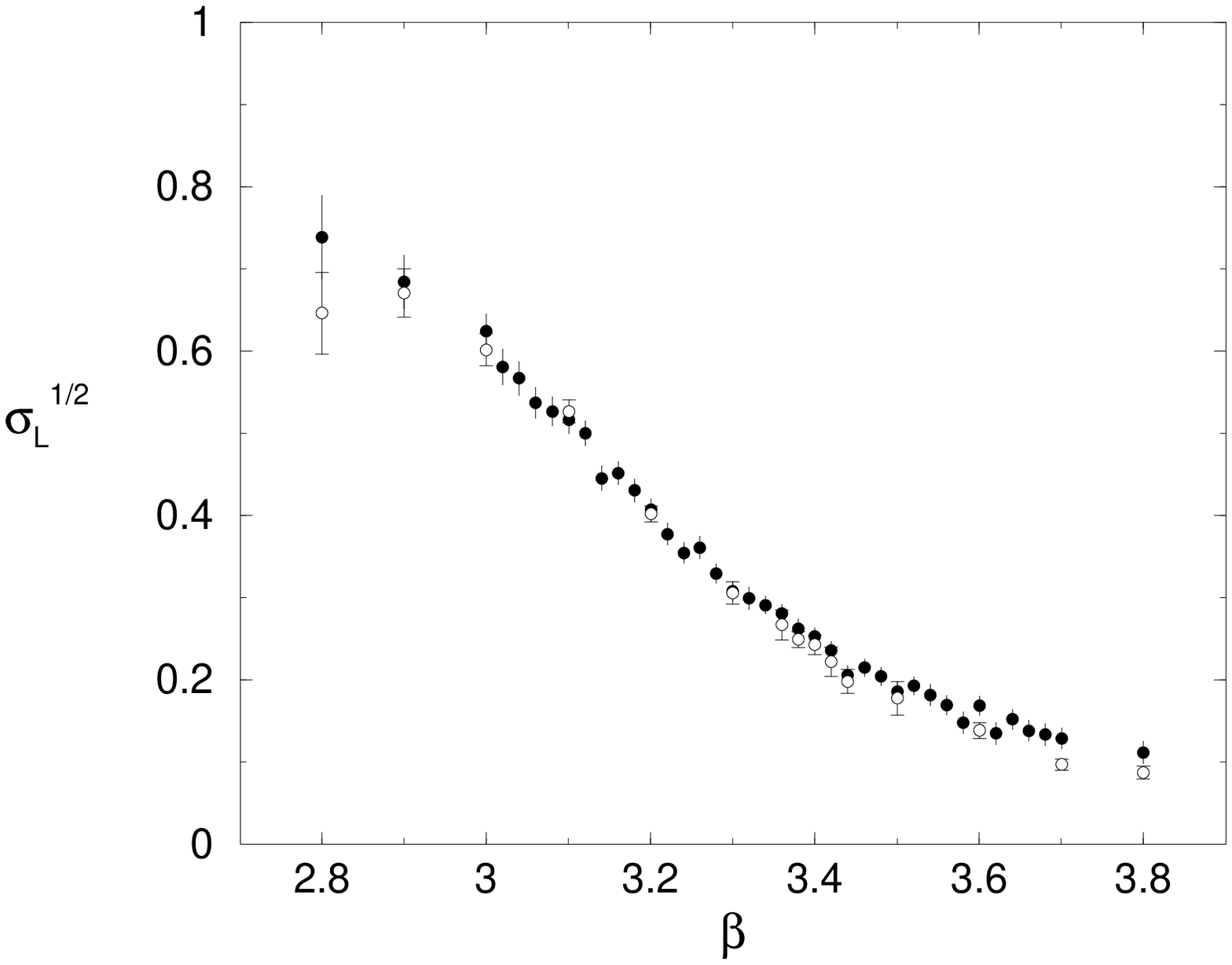}}
\caption{
$\sqrt{\chi_0} = \sqrt{\sigma_{L}}$ versus $\beta $ at 
$\gamma=5.0$. The filled symbols are obtained
from the Creutz ratio (\ref{chi}) and the open ones
are obtained from the static potential (\ref{st-pot}).
}
\label{fig1}
\end{figure}

\begin{figure}
\epsfxsize=12cm
\centerline{
\epsfbox{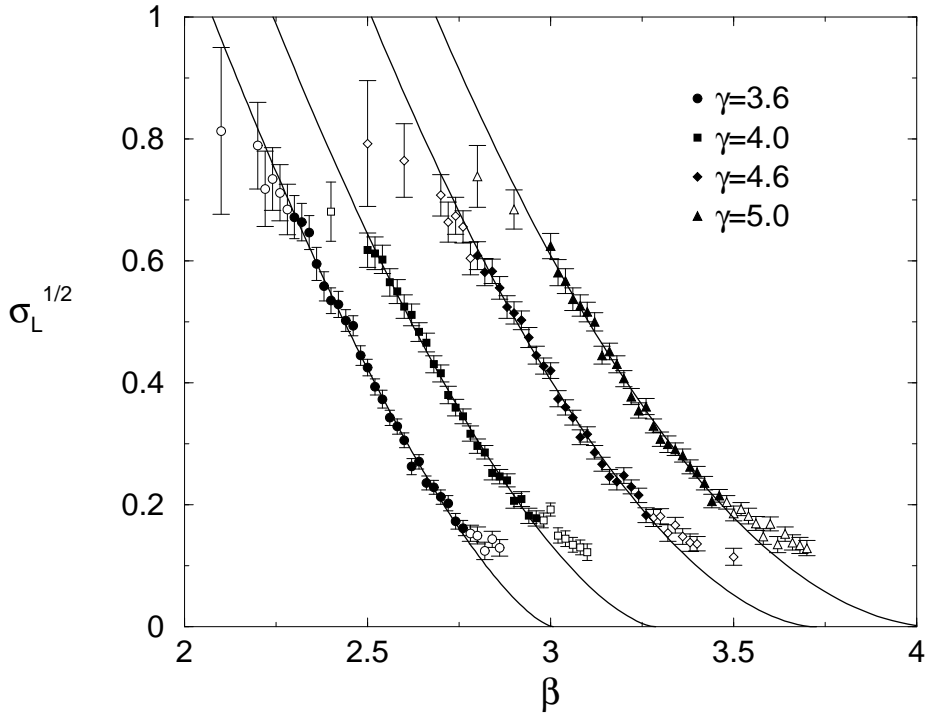}
}
\caption{The scaling behavior of $\sigma_L^{1/2}$
for different values of $\gamma$.
The solid lines are drawn by using (\ref{deff}), where
$D_{\rm eff}$ is taken from Table I.
The data with a filled symbol
are used for the fit.
}
\label{fig2}
\end{figure}

\begin{figure}
\epsfxsize=12cm
\centerline{
\epsfbox{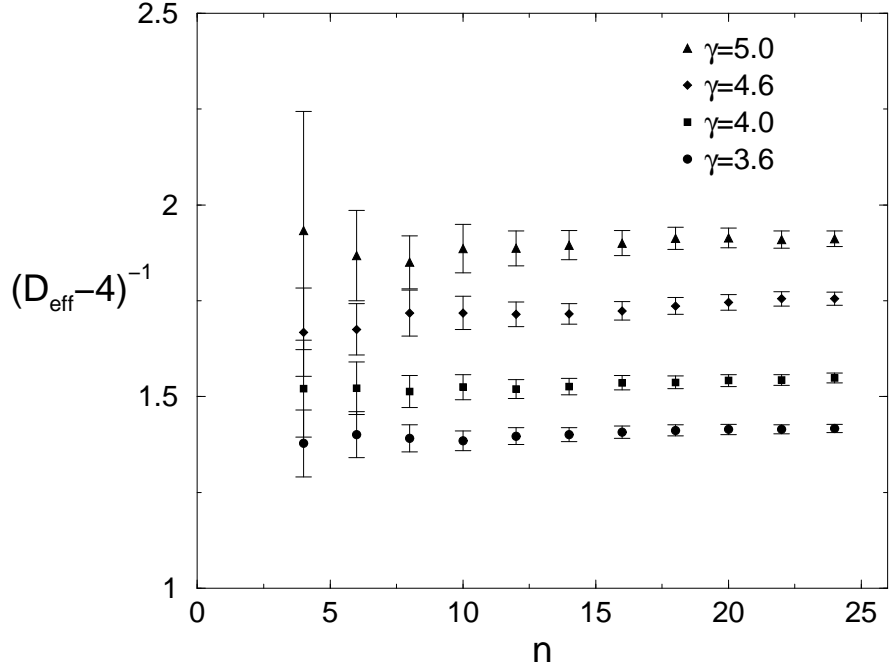}
}
\caption{
The effective dimension as a function of 
the number  $n$ of the data points that 
are used for  a fit. We increase $n$ 
starting from four till the value of $(D_{\rm eff}-4)^{-1}$ 
becomes stabilized. 
}
\label{fig3}
\end{figure}

\begin{figure}
\epsfxsize=12cm
\centerline{
\epsfbox{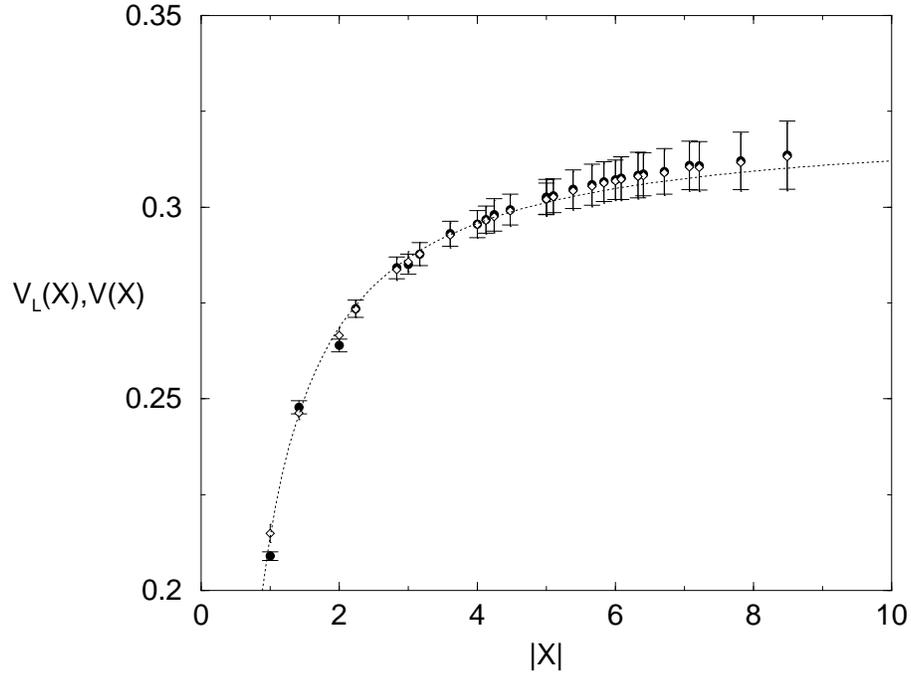}
}
\caption{The Coulomb potential (\ref{VL}).
The filled symbols are the raw data points, and the dotted line
is $V(X)$ of Eq. (\ref{pot1}) with 
$C_0=0.3230(14)  $ and $C_1=0.1086(30) $.
The open symbols stand for the 
rotationally invariant data points. 
}
\label{fig4}
\end{figure}

\begin{figure}
\epsfxsize=12cm
\centerline{
\epsfbox{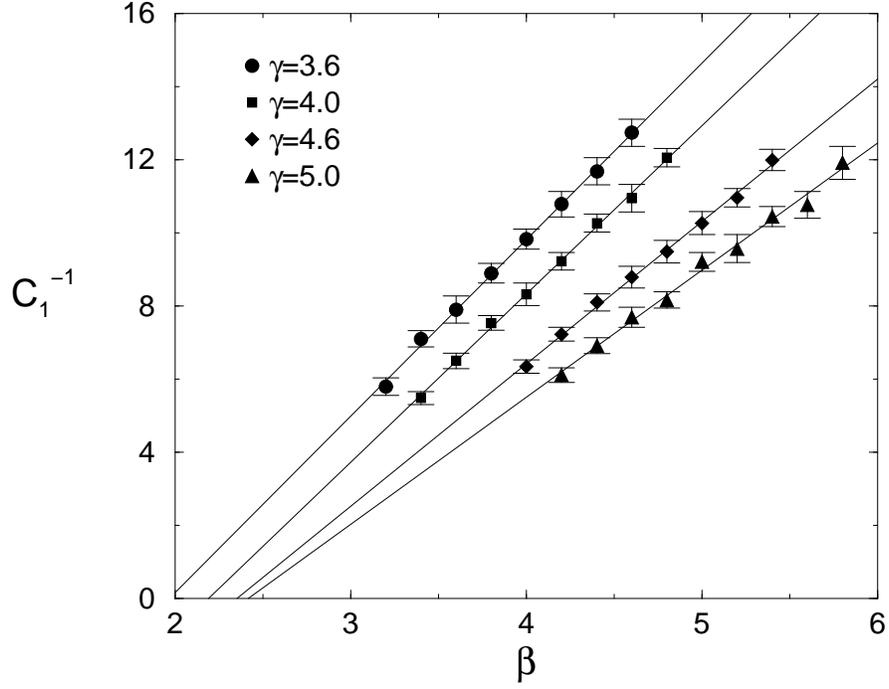}
}
\caption{$C_1^{-1}$ versus $\beta$ for
different $\gamma$'s, where $C_1^{-1}$
is defined in Eq. (\ref{pot}).
The graph shows the scaling behavior in the
Coulomb phase. The lines correspond to the 
linear function (\ref{c1fit}),
where $D_0$ and $D_1$ are given in Table II.
}
\label{fig5}
\end{figure}

\begin{table}
\caption{Effective dimension 
for different values of $\gamma$. ($\Lambda=2\pi/a_4$)}
\begin{tabular}{cclllll}
$\gamma$&  $R\Lambda$  & $D_{eff}$ &
$(\beta_{\mbox{\tiny min}}:\beta_{\mbox{\tiny max}})$ &
  $\chi^2/\mbox{DOF}$ & $\beta^*$ \\ \hline
3.6  & 0.72 & 4.7057(55) & (2.30 : 2.76) & 0.525 & 3.007(23) \\
4.0  & 0.64 &  4.6456(54) & (2.50 : 2.96) & 0.438 & 3.286(22) \\
4.6  &  0.55 & 4.5695(55) & (2.80 : 3.26) & 0.778 & 3.726(36) \\
5.0  & 0.50 &  4.5230(82) & (3.00 : 3.46) & 0.598 & 4.057(43) \\
\end{tabular}
\end{table}

\begin{table}
\caption{Fit for $C_1$ defined in (\ref{c1fit}).
The fitted lines in Fig.~5 intersect with
the $\beta$ axis at  $\beta=D_0/D_1$.}
\begin{tabular}{clllll}
$\gamma$  & $D_0$ & $D_1$ & 
$(\beta_{\mbox{\tiny min}} : \beta_{\mbox{\tiny max}})$ &
  $\chi^2/\mbox{DOF}$ & $D_0/D_1$ \\ \hline
3.6  &  9.48(31)  & 4.827(80) & (3.20 : 4.60) & 0.103  & 1.965(97) \\
4.0  & 10.08(30)  & 4.603(74) & (3.40 : 4.80) & 0.0957 & 2.19(10)  \\
4.6  &  9.16(36)  & 3.894(77) & (4.00 : 5.40) & 0.0998 & 2.35(14)  \\
5.0  &  8.39(57)  & 3.47(11)  & (4.20 : 5.80) & 0.175  & 2.41(24)  \\
\end{tabular}
\end{table}

\begin{table}
\caption{$\gamma$ independence of $\alpha$. }
\begin{tabular}{cll}
$\gamma$  & $\alpha$ \\ \hline
3.6 & 5.03(66) \\
4.0 & 5.05(64) \\
4.6 & 5.35(78) \\
5.0 & 5.7(11)  \\
\end{tabular}
\end{table}


\begin{references}

    \bibitem{kaluza}
Th.~Kaluza, Sitzungsber. d. Preuss. Akad. d. Wiss.,  966 (1921);
O.~Klein, Zeitschrift f. Phys. {\bf 37,} 895 (1926).

\bibitem{antoniadis1}
I.~Antoniadis, Phys. Lett. B {\bf 246,} 377 (1990); 
I.~Antoniadis, C.~Mu\~noz and M. ~\'os, Nucl. Phys. {\bf B397,} 515 (1993).

\bibitem{arkani1}
N.~Arkani-Hamed, S.~Dimopoulos and G.~Dvali,
Phys. Lett. B {\bf 429,} 263 (1998);
Phys. Rev. D {\bf 59,}  086004 (1999).

\bibitem{dienes1}
K.~Dienes, E.~Dudas and T.~Gherghetta,
Phys. Lett. B {\bf 436,} 55 (1998);
Nucl. Phys. {\bf B537,} 47 (1999).


\bibitem{creutz1}M.~Creutz, Phys. Rev. Lett. {\bf 43,} 553 (1979).

\bibitem{lang}
C.B.~ Lang, M.~Pilch and B.-S.~Skagerstam,
Int. J. Mod. Phys. A {\bf 3,} 1423 (1988).


\bibitem{okamoto1}H.~Kawai, M.~Nio and Y.~Okamoto,
Prog. Theor. Phys. {\bf 88,} 341 (1992);
J.~Nishimura, Mod. Phys. Lett. A {\bf 11,} 3049 (1996).

\bibitem{ejiri1}
S.~Ejiri, J.~Kubo and M.~Murata,
Phys. Rev. D {\bf D62,} 105025 (2000).

\bibitem{peskin} M.~Peskin, Phys. Lett. {\bf 94B,} 161 (1980).

\bibitem{veneziano}
T.R.~Taylor and G.~Veneziano, Phys. Lett. B {\bf 212,} 147 (1988).

\bibitem{kobayashi}
T.~Kobayashi, J.~Kubo, M.~Mondragon and  G.~Zoupanos,
Nucl. Phys. {\bf B550,} 99 (1999).

\bibitem{connor}
D.~O'Connor and C.~R.~Stephens, Phys. Rev. Lett. {\bf 72,} 
506 (1994); Int. J. Mod. Phys. {\bf A9,} 2805 (1994).

\bibitem{kubo2}
J.~Kubo, H.~Terao and G.~Zoupanos,
Nucl. Phys. {\bf B574,} 495 (2000).

\bibitem{arkani2}
N.~Arkani-Hamed, A.G.~Cohen and
H.~Georgi, Phys. Rev. Lett. {\bf 86,} 4757 (2001);
C.T~Hill, S.~Pokorski and J.~Wang,
Phys. Rev. D {\bf  64,} 105005 (2001).


\bibitem{burgers} 
G.~Burgers, F.~Karsch, A.~Nakamura and I.O.~Stamatescu, 
Nucl. Phys. {\bf B304,} 587 (1988);
T.R.~Klassen, Nucl. Phys. {\bf B533,} 557 (1998);
S.~Ejiri, Y.~Iwasaki, and K.~Kanaya,
Phys.\ Rev.\ D {\bf 58,} 094505 (1998);
J.~Engels, F.~Karsch and T.~Scheideler, 
Nucl.~Phys.~{\bf B564,} 303 (2000).

\bibitem{ape}
APE Collaboration, M.~Albanese {\it et al.},
Phys. Lett. B {\bf 192,} 163 (1987);
G.S.~Bali and K.~Schilling, Phys. Rev. D {\bf 47,} 661 (1993).

\bibitem{kubo3}
J.~Kubo and M.~Nunami,
hep-th/0112032.

\bibitem{murata}
M.~Murata, Ph.D. thesis, Kanazawa University, 
January 2002; M.~Murata and H.~So,
to appear.

\bibitem{fu}
Y. K.~ Fu and H.B.~ Nielsen,
Nucl. Phys. {\bf B236,} 167 (1984);
P.~Dimopoulos, K.~Farakos, A.~Kehagias and G.~Koutsoumbas,
Nucl. Phys. {\bf B617,} 237 (2001);
 P.~Dimopoulos, K.~Farakos and C.P.~Korthals-Altes,
JHEP 0102, 005 (2001).
P.~Dimopoulos, K.~Farakos  and S.~Nicolis,
hep-lat/0105014;
P.~Dimopoulos, K.~Farakos, G.~Koutsoumbas,
Phys. Rev. D {\bf 65,} 074505 (2002).


\bibitem{csaki}
C.~Cs\' aki, J.~Erlich, C.~Grojean
and G.D.~Kribs,
Phys. Rev. D {\bf 65,} 015003 (2001);
C.~Cs\' aki {\em et al.}, hep-th/0110188.


\end{references}
\end{document}